\documentclass[prd,amsmath,superscriptaddress,preprint]{revtex4-1}
\usepackage{amsmath}
\usepackage{amsfonts}
\usepackage{amssymb}
\usepackage{amsthm}

\usepackage{color}
\usepackage[normalem]{ulem}
 
\DeclareMathOperator{\arcsinh}{arcsinh}
 
\usepackage{graphicx}

\begin{document}

\preprint{\today}

\title{Standing wave solutions in Born-Infeld theory}

\author{Nenad Manojlovic}
\affiliation{Universidade do Algarve, Faculdade de Ci\^encias e Tecnologia, 8005-139 Faro, Portugal}
\affiliation{Grupo de F\'{\i}sica Matem\'atica da Universidade de Lisboa,
Campo Grande, Edif\'{\i}cio C6, 1749-016 Lisboa, Portugal}
\author{Volker Perlick}
\affiliation{ZARM, University of Bremen, Am Fallturm, 28159 Bremen, Germany}
\author{Robertus Potting}
\affiliation{Universidade do Algarve, Faculdade de Ci\^encias e Tecnologia, 8005-139 Faro, Portugal}
\affiliation{CENTRA, Instituto Superior T\'ecnico, Universidade de Lisboa,
Avenida Rovisco Pais, 1049-001 Lisboa, Portugal}

\begin{abstract}
We study standing-wave solutions of Born-Infeld electrodynamics,
with nonzero electromagnetic field in a region between two parallel 
conducting plates.
We consider the simplest case which occurs when the vector potential
describing the electromagnetic field has only one nonzero component
depending on time and on the coordinate perpendicular to the plates.
The problem then reduces to solving the scalar Born-Infeld equation,
a nonlinear partial differential equation in 1+1 dimensions.
We apply two alternative methods to obtain standing-wave solutions to
the Born-Infeld equation: an iterative method,
and a ``minimal surface'' method.
We also study standing wave solutions in a uniform constant magnetic
field background.

\end{abstract}

\maketitle
\vfill\pagebreak

\section{Introduction}

In 1934 Born and Infeld \cite{BornInfeld:1934} introduced a model of
nonlinear electrodynamics with the main objective to formulate an
alternative to Maxwell theory
with the property that pointlike charges have a finite self-energy.
The model is controlled by one parameter $b$ which is essentially the
maximum value of any purely electrostatic field configuration.
Another nonlinear modification of Maxwell theory was derived from
quantum electrodynamics by Euler and Heisenberg \cite{EulerHeisenberg:1936}.
Pleba\'nski \cite{Plebanski:1970} showed these models are two examples
of a large class of nonlinear electrodynamics (NLED) theories that are defined
by a Lorentz-invariant Lagrangian.
Nevertheless, Born-Infeld theory maintains a special place among
models of nonlinear electrodynamics as it proved to be the sole theory
with causal propagation \cite{Plebanski:1970,DeserPuzalowski:1980}
and absence of birefringence \cite{Boillat:1970} and is for this reason
called exceptional \cite{BialynickiBirula:1984}.
Renewed interest in the Born-Infeld theory has arisen in the 1980's
when it was shown to emerge in the low-energy limit of string theory
\cite{FradkinTseytlin:1985}.

In this paper we study standing-wave solutions of Born-Infeld electrodynamics,
where we assume the electromagnetic field to reside in a region
between two parallel conducting plates.
The simplest situation occurs when the electric and the magnetic field
are mutually orthogonal and parallel to the plates,
and when both only depend on time and on the coordinate perpendicular
to the plates.
In that case the equation of motion for the only nonzero component
of the vector potential reduces to the scalar Born-Infeld equation,
a nonlinear partial differential equation in 1+1 dimensions
\cite{Barbashov:1966frq}.
We review two methods to obtain its standing-wave solutions.
An iterative method that goes back to Poincar\'e and Lindstedt,
first applied by Ferraro \cite{Ferraro} to Born-Infeld theory,
starts from a seed function.
Repeated iterations yield corrections terms proportional to 
increasing powers of the inverse of the Born-Infeld parameter $b$,
resulting in an asymptotic series approximation to a standing-wave solution
of the Born-Infeld equation.
An alternative ``minimal surface'' method,
first developed by Barbaskov and Chernikov \cite{Barbashov:1966frq},
is based on the fact that the scalar Born-Infeld equation in 1+1 dimensions
is integrable.
We show how its solutions can be obtained in parametric form.
We also study standing wave solutions in a uniform constant magnetic
field background.

This paper is organized as follows.
In section \ref{sec:BI-equation} we show how the scalar Born-Infeld
equation can be obtained from the Lagrangian of Born-Infeld electrodynamics.
In section \ref{sec:iterative} we apply an iterative procedure to obtain
an approximation to the standing wave solutions of the Born-Infeld equation
in the form of an asymptotic series.
In section \ref{sec:minimal-surfaces} the minimal surface method
for obtaining exact solutions is reviewed
and applied in order to obtain standing-wave solutions to the Born-Infeld
equation with arbitrary initial conditions.
Conclusions and a discussion of the results is presented in
section \ref{sec:discussion}.

In this work we adopt the metric convention $(+,-,-,-)$ and use natural,
Heaviside-Lorentz units (with $c = \hbar = 1$).

\section{Born-Infeld theory and the scalar Born-infeld equation}
\label{sec:BI-equation}

We start with the Born-Infeld action
\begin{equation}
\mathcal{L} = b^2 - b^2\sqrt{1 + \frac{2F}{b^2} - \frac{G^2}{b^4}}
\label{BI-action}
\end{equation}
where $b$ is a fixed parameter of mass dimension 2, while
\begin{equation}
F = \frac{1}{4}F^{\mu\nu}F_{\mu\nu} \>, \qquad
G = \frac{1}{4}F^{\mu\nu}\tilde F_{\mu\nu}\>,
\end{equation}
with the electromagnetic field strength
\begin{equation}
F_{\mu\nu} = \partial_\mu A_\nu - \partial_\nu A_\mu
\end{equation}
and its dual
\begin{equation}
\tilde{F}^{\mu\nu} = \frac{1}{2}\epsilon^{\mu\nu\rho\sigma}F_{\rho\sigma}\>.
\end{equation}
The equations of motion resulting from Eq.\ \eqref{BI-action} are
analogous to the (source-free) Maxwell equations:
\begin{align}
\label{Ampere}
\partial_\nu(\mathcal{L}_F F^{\mu\nu} + \mathcal{L}_G\tilde{F}^{\mu\nu}) &= 0 \\
\partial_\nu \tilde F^{\mu\nu} &= 0
\label{Faraday}
\end{align}
where
\begin{equation}
\mathcal{L}_F = \frac{-1}{\sqrt{1 + \frac{2F}{b^2} - \frac{G^2}{b^4}}}
\qquad\mbox{and}\qquad
\mathcal{L}_G = \frac{\frac{G}{b^2}}{\sqrt{1 + \frac{2F}{b^2} - \frac{G^2}{b^4}}}
\end{equation}
In this work we will consider a strongly reduced subset of field configurations,
namely those in which the field components only depend on time and on one space
coordinate, which we will take to be $x$.
Moreover, we will only consider field configurations in which both the
electric and magnetic fields are linearly polarized perpendicular to $x$,
and are perpendicular to each other.
Without loss of generality, we can take $\vec{E}$ in the $y$ direction,
and $\vec{B}$ in the $z$ direction.
As is shown in appendix \ref{app:gauge-choice},
for our reduced subset of field configurations it is possible
to make a choice of gauge in which the gauge field takes the simple form
\begin{equation}
A_y(x,y,z,t) = u(x,t)\>,\qquad \Phi = A_x = A_z = 0\>.
\label{gauge-choice}
\end{equation}
for some field $u(x,t)$.
We then have
\begin{align}
F_{02} &= E_y = -\partial_t A_y = -u_t  \\
-F_{12} &= B_z = \partial_x A_y = u_x\>.
\end{align}

It is easy to verify that $G = \frac{1}{4}F_{\mu\nu} \tilde{F}^{\mu\nu} = 0$,
considerably simplifying the expressions.
Working out the equation of motion \eqref{Ampere},
it is not hard to show that it reduces to the following
nonlinear partial differential equation for $u(x,t)$
\begin{equation}
\left(1 - \frac{1}{b^2}u_t^2\right) u_{xx}
- \left(1 + \frac{1}{b^2}u_x^2\right) u_{tt}
+ \frac{2}{b^2}u_x u_t u_{xt} = 0
\label{BI-equation-b}
\end{equation}
which is known as the scalar Born-Infeld equation
(lower indices on $u$ indicate partial derivatives).

In this work we will take the Born-Infeld equation subject to Dirichlet
boundary conditions on $u$ at two parallel surfaces defined by, say,
$x = 0$ and $x = L$ and search for standing-wave-type oscillatory solutions.
Physically this is equivalent to placing two plane conducting
plates parallel to the $yz$ plane,
forcing the electric field to vanish at $x = 0$ and $x = L$.

For Eq.\ \eqref{BI-equation-b} to yield oscillatory solutions,
it should be in the hyperbolic regime. 
A necessary and sufficient conditon for this to be the case is
\begin{equation}
1 + \frac{1}{b^2}(u_x^2 - u_t^2) > 0\>,
\label{hyperbolic-regime}
\end{equation}
as one can easily verify.
Condition \eqref{hyperbolic-regime} will constrain the solutions
we will derive in sections \ref{sec:iterative} and \ref{sec:minimal-surfaces}.

\section{Standing wave solution of Born-Infeld electrodynamics: iterative method}
\label{sec:iterative}

Ferraro \cite{Ferraro} obtained an approximation to a standing wave solution
to the Born-Infeld equation \eqref{BI-equation-b} by writing first
Eq.\ \eqref{BI-equation-b} in the form
\begin{equation}
u_{xx} - u_{tt} - b^{-2}\bigl(u_t^2u_{xx} + u_x^2u_{tt} - 2 u_xu_tu_{xt}\bigr) = 0\>.
\label{BI-equation-iter}
\end{equation}
He took then as a seed a standing wave solution $u^{(0)}(x,t)$
to the linear part of Eq.\ \eqref{BI-equation-iter}
(i.e., the $b\to\infty$ limit of Eq.\ \eqref{BI-equation-iter},
corresponding to the usual Maxwell equations).
Being a standing wave solution,
$u^{(0)}(x,t)$ is periodic with characteristic (angular) frequency $\omega$.
Substituting $u(x,t) = u^{(0)}(x,t)$ and evaluating the full left-hand side of
Eq.\ \eqref{BI-equation-iter} yields a nonzero expression.
It can be shown that for it to vanish up to $\mathcal{O}(b^{-2})$
it is necessary to add an $\mathcal{O}(b^{-2})$ correction $u^{(1)}$
to $u^{(0)}$, which consists of a periodic part and a nonperiodic part.
As it turns out, the latter can be absorbed into $u^{(0)}(x,t)$
by adding an $\mathcal{O}(b^{-2})$ correction to the value of the frequency.
The resulting sum $u = u^{(0)} + u^{(1)}$ is then periodic in time and
satisfies Eq.\ \eqref{BI-equation-iter} to order $b^{-2}$.
This procedure is referred to in the literature as the Poincar\'e-Lindstedt method \cite{Lindstedt:1882,Poincare:1957}.

\subsection{Standing wave solutions}

We will now review this procedure in detail,
and show how the solution can be extended to higher orders $\mathcal{O}(b^{-2n})$.
We start with the seed solution
\begin{equation}
u^{(0)} = A \sin k x \cos\omega t = 
\frac{A}{2}\bigl(\sin(k x + \omega t) +\sin(k x - \omega t)\bigr)
= \frac{A}{2}s_{11}(x,t)
\label{eq:seed}
\end{equation}
where we introduced the notation
\begin{equation}
s_{nm}(x,t) = \sin(nk x + m\omega t) +\sin(nk x - m\omega t)\>.
\end{equation}
It satisfies the linear part of Eq.\ \eqref{BI-equation-iter}
by taking $\omega^2 = k^2$.
Taking now $u = u^{(0)}$ and evaluating the left-hand side of
\eqref{BI-equation-iter}, one obtains
\begin{equation}
\frac{A}{2}(\omega^2 - k^2)s_{11} - \frac{A^3k^2\omega^2}{8b^2}(s_{13} - s_{31} - 2s_{11})\>.
\label{u0-eq}
\end{equation}
In order to cancel the terms in Eq.\ \eqref{u0-eq} to order $b^{-2}$,
first of all we need to make the coefficient of $s_{11}$ to vanish.
It follows that
\begin{equation}
\omega^2 - k^2 + \frac{\omega^2\epsilon^2}{2} = 0\>,
\label{eq-omega-1}
\end{equation}
in which we defined the dimensionless quantity
\begin{equation}
\epsilon = \frac{A k}{b}\>.
\label{epsilon-1}
\end{equation}
This yields a corrected value for the frequency:
\begin{equation}
\frac{\omega^2}{k^2} = 1 - \frac{\epsilon^2}{2} + \mathcal{O}(\epsilon^4)
\label{omega-1}
\end{equation}

Secondly, we need to eliminate to terms proportional to $s_{13}$ and $s_{31}$.
Noting that
\begin{equation}
\left(\frac{\partial^2}{\partial x^2} - \frac{\partial^2}{\partial t^2}\right)
s_{nm} = (m^2\omega^2 - n^2 k^2)s_{nm} 
\end{equation}
it follows that this can be done at this order by adding to $u$ the
$\mathcal{O}\bigl(\epsilon^2\bigr)$ contribution
\begin{align}
u^{(1)} &= \frac{A\epsilon^2}{8}\left(\frac{s_{13}}{9\omega^2 - k^2} - \frac{s_{31}}{\omega^2 - 9k^2}\right) \nonumber\\
&= \frac{A\epsilon^2}{64}\left(s_{13} + s_{31}\right) + \mathcal{O}\bigl(\epsilon^4\bigr)\>.
\label{u1}
\end{align}
Eqs.\ \eqref{omega-1} and \eqref{u1} confirm the result obtained in
\cite{Ferraro}.

Let us now continue to the next order.
Evaluating the left-hand side of Eq.\ \eqref{BI-equation-iter} for
$u = u^{(0)} + u^{(1)}$ to $\mathcal{O}(\epsilon^4)$ one finds
a linear combination of the functions $s_{nm}(x,t)$,
with $n$ and $m$ taking values up to 5.
Considering first the coefficient of the seed function $s_{11}$,
it turns out that there is no contribution to Eq.\ \eqref{eq-omega-1} at
$\mathcal{O}(\epsilon^4)$.
It therefore follows that
\begin{equation}
\frac{\omega^2}{k^2} = 1 - \frac{\epsilon^2}{2} + \frac{\epsilon^4}{4}
+ \mathcal{O}(\epsilon^6)\>.
\label{omega-2}
\end{equation}
Analogously to the procedure outlined above at order $\epsilon^2$,
the coefficients of the remaining functions $s_{nm}(x,t)$ can be
made to cancel by adding to $u$ the
$\mathcal{O}\bigl(\epsilon^4\bigr)$ contribution
\begin{equation}
u^{(2)} = -\frac{A\epsilon^4}{1024}\left(5s_{13} + 3s_{31} - s_{15} - s_{51} + \frac{s_{35} - s_{53}}{4}\right)\>.
\label{u2}
\end{equation}

Substituting the corrected solution $u = u^{(0)} + u^{(1)} + u^{(2)}$
into the left-hand side of Eq.\ \eqref{BI-equation-iter} and
evaluating to $\mathcal{O}(\epsilon^6)$ one finds
a linear combination of the functions $s_{nm}(x,t)$,
where now $n$ and $m$ take odd values up to 7.
The coefficient of the seed function $s_{11}$ yields an additional
contribution $-3A\omega^2\epsilon^6/2048$ to the left-hand side of Eq.\ \eqref{eq-omega-1},
from which we deduce the corrected dispersion relation
\begin{equation}
\frac{\omega^2}{k^2} = 1 - \frac{\epsilon^2}{2} + \frac{\epsilon^4}{4}
- \frac{125\epsilon^6}{1024} + \mathcal{O}(\epsilon^8)\>.
\label{omega-3}
\end{equation}
The coefficients of the remaining functions $s_{nm}(x,t)$ can be
made to cancel by adding to $u$ the
$\mathcal{O}\bigl(\epsilon^6\bigr)$ contribution
\begin{align}
u^{(3)} &= \frac{A\epsilon^6}{32768}\biggl(\frac{189s_{13} + 61s_{31}}{4} - 17s_{15} - 15s_{51} + 3s_{35} - 5s_{53}
\nonumber\\
&\qquad\qquad\qquad{} + 3s_{17} + 3s_{71} - s_{37} + s_{73} + \frac{s_{57} + s_{75}}{12}\biggr)\>.
\label{u3}
\end{align}
Altogether, we have
\begin{align}
u &= \frac{A}{2}\Biggl[s_{11} + \frac{\epsilon^2}{32}\bigl(s_{13} + s_{31}\bigr) - \frac{\epsilon^4}{512}\left(5s_{13} + 3s_{31}
- s_{15} - s_{51} + \frac{s_{35} - s_{53}}{4}\right)\nonumber\\
&\quad\qquad{}+ \frac{\epsilon^6}{16384}\biggl(\frac{189s_{13} + 61s_{31}}{4} - 17s_{15} - 15s_{51} + 3s_{35} - 5s_{53}
\nonumber\\
&\qquad\qquad\qquad\qquad{} + 3s_{17} + 3s_{71} - s_{37} + s_{73} + \frac{s_{57} + s_{75}}{12}\biggr) + \mathcal{O}(\epsilon^8)\Biggr]\>.
\label{u-sixth-order}
\end{align}

This procedure can be continued to higher orders $\epsilon^{2N}$.
Explicitly, to solve the Born-Infeld equation to order $\epsilon^{2N}$,
we have $u = u^{(0)} + u^{(1)} + \ldots + u^{(N)}$,
where $u^{(k)}$ involves contributions to coefficient functions
$s_{ij}$ with $1\le i,j \le 2k+1$, $i$ and $j$ odd, at order $\epsilon^{2k}$.
For the dispersion relation we find a power series
\begin{equation}
\frac{\omega^2}{k^2} = 1 - \frac{\epsilon^2}{2} + \frac{\epsilon^4}{4}
+ \ldots\>.
\label{omega-power series}
\end{equation}
where the coefficients of $\epsilon^{2k}$ are determined order by order.
This way,
the coefficients of the individual functions $s_{ij}$ for $i$ and $j$ not
both equal to 1 become a power series in $\epsilon$.
It is interesting to speculate these series are actually convergent
(absolutely or pointwise). We have worked out the general recursion 
formulas for the coefficients of the series, see Appendix \ref{app:asymptotic},
and with the help of these formulas we have tried to prove, or disprove, 
convergence.  Unfortunately we have not been able to do so. 

However, it is easy to see that
$u(x,t)$ does satisfy an alternative criterion.
Let us write the Born-Infeld equation \eqref{BI-equation-iter} as
$D_b u  = 0$, where the nonlinear differential operator $D_b$ depends on $b$.
We define $u(x,t)$ to be an \textit{asymptotic solution of order} $N$
if $b^{2N} D_b u \to 0$ for $b \to \infty$ (pointwise, say).
It is evident from our iterative procedure that
it produces asymptotic solutions of arbitrarily high order $N$,
because after $N$ iterations the approximative solution
$u \approx u^{(0)} + u^{(1)} +\ldots + u^{(N)}$
yields an expression of order $b^{-2N-2}$ when we evaluate $D_b u$.
We conjecture that this solution approaches an exact
solution for $N \to \infty$. Asymptotic series solutions are often more
useful in view of approximations than slowly converging series. E.g., 
it is well known that the WKB series which are discussed in any
textbook on quantum mechanics are not in general convergent.
However, they often give very good approximations even if
cut off after only a few terms. We will numerically estimate
how accurately our asymptotic series solutions satisfy the 
Born-Infeld equations at the end of Appendix \ref{app:asymptotic}.

As noted at the end of section \ref{sec:BI-equation},
for the Born-Infeld equation to allow oscillatory solutions,
the solution $u(x,t)$ has to satisfy the hyperbolicity condition
\eqref{hyperbolic-regime} for all values of $x$ and $t$.
While this condition strictly only applies to the exact
solution to Eq.\ \eqref{BI-equation-iter},
it will be satisfied as well at least approximately
for the iterative approximations $u^{(N)}$,
with increasing precision for increasing values of $N$.
It is straightforward to verify that $u^{(0)}$ satisfies the hyperbolic 
condition \eqref{hyperbolic-regime} for any $x$ and $t$ if and only if
the maximum amplitude of the seed function \eqref{eq:seed} satisfies
$Ak < b$, or, equivalently, $\epsilon < 1$. 
This demonstrates the existence of a maximum critical value for $\epsilon$
in order for hyperbolicity to hold. 
While this value may shift for higher values of $N$,
it should remain of order unity.
It is interesting to note from Fig.\ \ref{fig:error},
which indicates the error with which the approximations satisfy the
Born-Infeld equation,
that at any level of approximation the error rises with $\epsilon$.
From the argument above, we can actually expect that the approximations
will fail to converge at all above some critical value for $\epsilon$
of order unity.

As a final point we remark that the solution Eq.\ \eqref{u-sixth-order}
satisfies the periodicity conditions
\begin{equation}
u\left(x + \frac{2\pi}{k},t\right) = u(x,t) = u\left(x,t + \frac{2\pi}{\omega}\right)
\end{equation}
and the antisymmetry condition
\begin{equation}
u(-x, t) = -u(x,t)
\end{equation}
Therefore, $u$ satisfies the (standing-wave) boundary conditions
\begin{equation}
u(0,t) = u(L,t) = 0
\end{equation}
for $L = \pi/k$.

\subsection{More general seed functions}

Let us now see how the method generalizes to the case in which the seed
function has a more complicated form. 
Ideally, we would like to take for the seed an arbitrary Fourier series
\begin{equation}
u^{(0)}(x,t) = \sum_{n=1}^\infty \sin(nkx) \bigl(A_n \cos(\omega_n t) + B_n \sin(\omega_n t)\bigr) 
\label{general-u0}
\end{equation}
satisfying the boundary conditions as well as the linearized equation of
motion for $\omega_n^2 = n^2k^2$.
Here we will be a bit less ambitious, and take
\begin{equation}
u^{(0)}(x,t) = A_1\sin(kx)\cos(\omega t) + A_3\sin(3kx)\cos(3\omega t)
= \frac{A_1}{2}s_{11}(x,t) + \frac{A_3}{2}s_{33}(x,t)\>.
\label{u0-two_terms}
\end{equation}
Note that we have identified $\omega_1 = \omega$, $\omega_3 = 3\omega$.
While this is obviously correct at order $b^0$, it is not evident that
the corrections to the frequency of the two modes $s_{11}$ and $s_{33}$
at higher orders will maintain this simple relation.
However, note that if the relation were to break down at higher orders,
it would mean that the nonlinear solution is not a periodic function
in time.
We will see in section \ref{sec:minimal-surfaces},
however, that this is in fact the case.

Taking $u = u^{(0)}$ and evaluating the left-hand side of Eq.\ \eqref{BI-equation-iter},
one obtains
\begin{align}
&\left(\omega^2\left(1 + \frac{\epsilon^2}{2}\right) - k^2\right)
\left(\frac{A_1}{2}s_{11} + \frac{9A_3}{2}s_{33}\right) \nonumber\\
&\quad{} + \frac{k^2\omega^2}{8b^2}\bigl[A_1^2(A_1 + 6A_3)(s_{31} - s_{13})
+ 9A_1 A_3(A_1 + 6A_3)(s_{51} + s_{15}) + 6A_1^2A_3(s_{53} - s_{35})\nonumber\\
&\quad\qquad\qquad{} + 54A_1 A_3^2(s_{71} - s_{17}) + 9A_1A_3^2(s_{75} - s_{57}) + 81A_3^3(s_{93} - s_{39})\bigr] \>,
\label{u0-eq-2}
\end{align}
where we defined
\begin{equation}
\epsilon = \frac{k\sqrt{A_1^2 + 9A_3^2}}{b} \>,
\label{epsilon-2}
\end{equation}
modifying the definition \eqref{epsilon-1}.
We see from expression \eqref{u0-eq-2} that the coefficients of the diagonal terms $s_{11}$ and $s_{33}$
can be made to vanish by imposing the dispersion relation \eqref{omega-1}
at this order, but with $\epsilon$ now given by relation \eqref{epsilon-2}.
Note that this confirms that the order $b^{-2}$ corrections to the frequencies
of the modes $s_{11}$ and $s_{33}$ are in fact identical,
as anticipated above. 

The off-diagonal terms in Eq.\ \eqref{u0-eq-2} can be canceled by
the order $b^{-2}$ correction
\begin{align}
u^{(1)} &= \frac{k^2}{8b^2}\biggl[\frac{A_1^2(A_1 + 6A_3)}{8}(s_{31} + s_{13})
+ \frac{9A_1 A_3(A_1 + 6A_3)}{24}(s_{51} - s_{15}) + \frac{3A_1^2A_3}{8}(s_{53} + s_{35})\nonumber\\
&\qquad\qquad{} + \frac{54A_1 A_3^2}{48}(s_{71} + s_{17}) + \frac{9A_1A_3^2}{24}(s_{75}
+ s_{57}) + \frac{9A_3^3}{8}(s_{93} + s_{39})\biggr]\>.
\label{u1-2}
\end{align}
Of course, taking $A_3 = 0$, $A_1 = A$, the correction \eqref{u1-2}
reduces to the case \eqref{u1}.

We thus conclude that the generalization of the iterative method to
seeds that are linear combinations of the modes $s_{nn}$ is
in principle straightforward.

\subsection{Standing waves in a magnetic field background}

Let us now try to generalize the above solution by assuming the presence
of a constant external magnetic field in the $z$-direction.
Thus we take as the seed for the solution
\begin{equation}
u^{(0)} = \frac{A}{2} s_{11}(x,t) + B.x
\end{equation}
At order $b^0$ nothing changes.

At order $b^{-2}$ we find
\begin{equation}
\bigl(u_t^{(0)}\bigr) u_{xx}^{(0)} + \bigl(u_x^{(0)}\bigr) u_{tt}^{(0)}
- 2u_x^{(0)} u_t^{(0)} u_{xt}^{(0)} =
\frac{A\epsilon^2}{2}\left[-\left(\frac{\omega^2}{4} + \frac{B^2}{A^2}\right)
s_{11} - \frac{\omega^2}{4}\bigl(s_{13} + s_{31}\bigr) - \frac{B}{A}\omega s_{20}\right]\>.
\end{equation}
Thus, the frequency suffers an additional shift proportional to the
square of $B$:
\begin{equation}
\frac{\omega^2}{k^2} = 1 - \left(\frac{1}{4} + \left(\frac{B}{Ak}\right)^2\right)
\epsilon^2 + \mathcal{O}(\epsilon^4)\>.
\label{dispersion_B-ext}
\end{equation}
Moreover, we need to add a contribution $u^{(1)}$ such that
\begin{equation}
u^{(1)}_{xx} - u^{(1)}_{tt} = \frac{A\epsilon^2\omega^2}{8}\bigl(s_{13} - s_{31}\bigr)
- \frac{B\omega\epsilon^2}{2}s_{20}\>.
\end{equation}
From this it follows that
\begin{equation}
u^{(1)} = \frac{\epsilon^2}{64}\left(A\bigl(s_{13} + s_{31}\bigr)
+ \frac{8B}{k}\,s_{20}\right) + \mathcal{O}(\epsilon^4)\>.
\end{equation}

It is worthwhile to point out that in the limit of large external field,
$B \gg Ak$, Eq.\ \eqref{dispersion_B-ext} yields
\begin{equation}
\frac{\omega}{k} \approx \sqrt{1 - \frac{B^2}{b^2}}
\approx 1 - \frac{B^2}{2b^2}
\label{dispersion_B-ext-limit}
\end{equation}
so that the frequency shift is determined by the external magnetic field.

\section{Standing wave solution of Born-Infeld electrodynamics: the method of minimal surfaces}
\label{sec:minimal-surfaces}

It is well known that the 1+1-dimensional Born-Infeld equation is an
integrable system.
Its general solution was obtained in parametric form by Barbashov and 
Chernikov \cite{Barbashov:1966frq}.
Later it was shown that the system has a multi-Hamiltonian structure
with associated conservation laws and higher symmetries \cite{NutkuOlver:89}.
A Lax representation can be obtained yielding all conserved charges
\cite{Brunelli:1997kh}.
The B\"acklund transformations for the Born-Infeld equation were first
introduced in \cite{NutkuOlver:89} and studied further in \cite{Menshikh2005}. 
These transformations can be used to generate new solutions starting
from the known ones \cite{Gutshabash:2017qxl}.
Some classes of exact solutions of the Born-Infeld equation were studied
in \cite{Mallory2014}. 

\subsection{The method of Barbashov and Chernikov}
In this section we will apply the method of Barbashov and Chernikov
to the case of standing wave solutions, allowing to obtain its general
solution in parametric form and study its properties.

We consider the Born-Infeld equation \eqref{BI-equation-b}
\begin{equation}
(1 - u_t^2)u_{xx} - 2u_x u_t u_{xt} - (1 + u_x^2)u_{tt} = 0
\label{BI-equation}
\end{equation}
in which 
we have chosen units such that the Born-Infeld parameter $b$ equals one,
in order to simplify the notation.

We first review the method by which Barbashov and Chernikov obtained
solutions of Eq.\ \eqref{BI-equation} for arbitrary initial conditions
\cite{Barbashov:1966frq} in the hyperbolic regime
\begin{equation}
1 + u_x^2 - u_t^2 > 0\>.
\label{hyperbolic-regime_no-b}
\end{equation}
The initial conditions are taken to be
\begin{align}
\label{initial-condition_a}
u|_{t=0} &= a(x)\>, \\
u_t|_{t=0} &= b(x)
\label{initial-condition_b}
\end{align}
where the functions $a(x)$ and $b(x)$ satisfy the hyperbolicity condition
\begin{equation}
1 + a'{}^2(x) - b^2(x) >
\label{hyperbolic_a,b}
\end{equation}
corresponding to Eq.\ \eqref{BI-equation}. Next,
new independent variables $\alpha$ and $\beta$ are introduced such that
\begin{equation}
x = x(\alpha,\beta)\>, \qquad t = t(\alpha,\beta)\>,
\qquad z(\alpha,\beta) = u\bigl(x(\alpha,\beta),t(\alpha,\beta)\bigr)\>.
\label{alpha-beta}
\end{equation}
It is convenient to consider $t$, $x$ and $z$ to be the components of a vector
$\vec{r}(\alpha,\beta)$ living in a Minkowskian space with scalar product
\begin{equation}
\vec{r}_1\cdot \vec{r}_2 = t_1 t_2 - x_1 x_2 - z_1 z_2 \>.
\end{equation}
The freedom in defining $\alpha$ and $\beta$ can now be used such that,
in the hyperbolic regime, $\vec{r}(\alpha,\beta)$ satisfies
\cite{Courant-Hilbert}
\begin{equation}
\vec{r}_{,\alpha}{}^2 = 0\>,\qquad \vec{r}_{,\beta}{}^2 = 0\>.
\label{characterics}
\end{equation}
The Born-Infeld equation \eqref{BI-equation} then reduces to the simple condition
\begin{equation}
\vec{r}_{,\alpha\beta} = 0\>.
\label{BI-equation2}
\end{equation}
The lower indices preceded by a comma in Eqs.\ \eqref{characterics}
and \eqref{BI-equation2} indicate partial derivatives with respect to
$\alpha$ and $\beta$.
The general solution to Eqs.\ \eqref{characterics} and \eqref{BI-equation2}
is clearly
\begin{equation}
\vec{r}(\alpha,\beta) = \vec{r}_1(\alpha) + \vec{r}_2(\beta)
\label{general-solution}
\end{equation}
where the vector functions  $\vec{r}_1$ and $\vec{r}_2$ satisfy
\begin{equation}
\vec{r}_1{}'{\,}^2(\alpha) = \vec{r}_2{}'{\,}^2(\beta) = 0
\label{lightlike-conditions}
\end{equation}
(the prime indicates derivative with respect to the argument).

Conditions \eqref{characterics} and \eqref{BI-equation2} actually only
determine the parameters $\alpha$ and $\beta$ up to an arbitrary
reparametrization
\begin{equation}
\alpha = A(\alpha')\>, \qquad \beta = B(\beta')
\end{equation}
where $A$ and $B$ are arbitrary monotonously increasing or decreasing functions.
Let us now express $\alpha$ and $\beta$ as functions of $x$ and $t$,
\begin{equation}
\alpha = \alpha(x,t)\>, \qquad \beta = \beta(x,t)\>,
\end{equation}
and set
\begin{equation}
A(x) = \alpha(x,0)\>, \qquad B(x) = \beta(x,0)\>.
\end{equation}
With this choice, the condition $t = 0$ is expressed in the new variables
$\alpha'$ and $\beta'$ as $\alpha' = \beta' = x$.
Omitting primes, initial condition \eqref{initial-condition_a} now implies:
\begin{align}
t(\alpha,\beta)|_{\beta=\alpha} &= t_1(\alpha) + t_2(\alpha) = 0 \nonumber\\
x(\alpha,\beta)|_{\beta=\alpha} &= x_1(\alpha) + x_2(\alpha) = \alpha \nonumber\\
z(\alpha,\beta)|_{\beta=\alpha} &= z_1(\alpha) + z_2(\alpha) = a(\alpha)\>,
\label{initial-condition_a2}
\end{align}
while initial condition \eqref{initial-condition_b} can be shown to yield
\begin{equation}
\left.\frac{\partial u}{\partial t}\right|_{\beta = \alpha} =
\frac{x_1'(\alpha)z_2'(\alpha) - z_1'(\alpha)x_2'(\alpha)}{x_1'(\alpha)t_2'(\alpha) - t_1'(\alpha)x_2'(\alpha)} = b(\alpha)\>,
\label{initial-condition_b2}
\end{equation}
by expressing the partial derivative with respect to $t$
in terms of partial derivatives with respect to $\alpha$ and $\beta$.

It is now convenient to split the general solution \eqref{general-solution}
into the sum of terms that are symmetric and antisymmetric in $\alpha$
and $\beta$ and write it in the form
\begin{equation}
\vec{r}(\alpha,\beta) = \frac{1}{2}\bigl(\vec{\rho}(\alpha) + \vec{\rho}(\beta)\bigr) + \frac{1}{2}\int_\alpha^\beta \vec{\pi}(\lambda)d\lambda\>.
\end{equation}
It then follows from condition \eqref{initial-condition_a2} that
\begin{equation}
\vec{\rho}(\alpha) = \bigl(0, \alpha, a(\alpha)\bigr)\>,
\end{equation}
while the vector
$\vec{\pi}(\lambda) = \bigl(\pi_t(\lambda),\pi_x(\lambda),\pi_z(\lambda)\bigr)$
is determined from conditions \eqref{lightlike-conditions} and
\eqref{initial-condition_b2}:
\begin{align}
\label{pi_t}
\pi_t(\lambda) &= \frac{1 + a'{}^2(\lambda)}{\sqrt{1 + a'{}^2(\lambda) - b^2(\lambda)}} \\
\pi_x(\lambda) &= \frac{-a'{}(\lambda)b(\lambda)}{\sqrt{1 + a'{}^2(\lambda) - b^2(\lambda)}} \\
\pi_z(\lambda) &= \frac{b(\lambda)}{\sqrt{1 + a'{}^2(\lambda) - b^2(\lambda)}}
\end{align}
It is interesting to note that $\pi_t$ is the Hamiltonian density,
$\pi_x$ the momentum density and $\pi_z$ the canonical momentum
of the field $u(x,t)$ at $t = 0$.

Concluding, the solution to the Born-Infeld equation \eqref{BI-equation} satisfying the initial conditions \eqref{initial-condition_a} and 
\eqref{initial-condition_b} has the parametric form:
\begin{align}
\label{sol-t}
t(\alpha,\beta) &= \frac{1}{2}\int_\alpha^\beta\frac{1 + a'{}^2(\lambda)}{\sqrt{1 + a'{}^2(\lambda) - b^2(\lambda)}} d\lambda \\
\label{sol-x}
x(\alpha,\beta) &= \frac{\alpha + \beta}{2} - \frac{1}{2}\int_\alpha^\beta\frac{a'{}(\lambda)b(\lambda)}{\sqrt{1 + a'{}^2(\lambda) - b^2(\lambda)}} d\lambda \\
\label{sol-z}
z(\alpha,\beta) &= \frac{a(\alpha) + a(\beta)}{2} + \frac{1}{2}\int_\alpha^\beta \frac{b(\lambda)}{\sqrt{1 + a'{}^2(\lambda) - b^2(\lambda)}} d\lambda\>.
\end{align}

\subsection{Standing wave solutions}

We will now search for standing wave solutions of the Born-Infeld equation
\eqref{BI-equation}.
We define the standing wave in the interval $x\in [0,L]$,
with Dirichlet boundary conditions
\begin{equation}
u(0,t) = u(L,t) = 0\>,
\label{boundary-conditions_u}
\end{equation}
implying that
\begin{equation}
a(0) = a(L) = b(0) = b(L) = 0\>.
\label{boundary-conditions_a,b}
\end{equation}
Conditions \eqref{boundary-conditions_a,b} can be satisfied by taking
$a(x)$ and $b(x)$ to be defined for any real value of $x$,
subject to the conditions
\begin{equation}
a(-x) = -a(x)\>,\qquad b(-x) = -b(x)\>,
\label{bc-0}
\end{equation}
as well as
\begin{equation}
a(L - x) = -a(L + x)\>,\qquad
b(L - x) = -b(L + x)\>.
\label{bc-pi}
\end{equation}
Combining conditions \eqref{bc-0} and \eqref{bc-pi},
it follows that $a$ and $b$ are periodic:
\begin{equation}
a(x + 2L) = a(x)\>,\qquad
b(x + 2L) = b(x)\>.
\label{bc-periodic}
\end{equation}
It is not hard to see that, if $a(x)$ and $b(x)$ satisfy
the conditions \eqref{bc-0} and \eqref{bc-periodic},
the boundary conditions \eqref{boundary-conditions_u} are satisfied
for all $t$.

To this effect, we first note that under the discrete transformation
$\alpha \to -\beta$, $\beta \to -\alpha$, 
the solutions \eqref{sol-t}, \eqref{sol-x} and \eqref{sol-z} transform as
\begin{equation}
t \to t\>, \qquad x \to -x\>, \qquad z \to -z\>.
\end{equation}
Thus it follows that
\begin{equation}
u(-x,t) = z(-\alpha,-\beta) = -z(\alpha,\beta) = -u(x,t)
\label{u-antisymmetry}
\end{equation}
so that $u$ satisfies the first condition in Eq.\ \eqref{boundary-conditions_u}
for all $t$.

Next, it is easy to see that under the transformation
$\alpha \to \alpha + 2L$, $\beta \to \beta + 2L$ we have
\begin{equation}
t \to t\>,\qquad x \to x + 2L\>,\qquad z \to z\>.
\end{equation}
Therefore
\begin{equation}
u(x+2L,t) = z(\alpha + 2L,\beta + 2L) = z(\alpha,\beta) = u(x,t)\>,
\label{u-x-periodicity}
\end{equation}
so that $u(x,t)$ is periodic in $x$ with period $2L$ for any $t$.
Combining Eqs.\ \eqref{u-antisymmetry} and \eqref{u-x-periodicity}
for $x = -L$,
the second condition in Eq.\ \eqref{boundary-conditions_u}
now follows directly for all $t$.

Finally, under the transformation $\alpha \to \alpha - 2L$,
$\beta \to \beta + 2L$ we have
\begin{equation}
t \to t + 2K\>,\qquad x \to x\>,\qquad z \to z\>,
\end{equation}
where the constant $K$ is defined by
\begin{equation}
K = \frac{1}{2}\int_0^{2L}\frac{1 + a'{}^2(\lambda)}{\sqrt{1 + a'{}^2(\lambda) - b^2(\lambda)}} d\lambda \ge L\>,
\label{K}
\end{equation}
where $K = L$ only in the trivial case $a(\lambda) = b(\lambda) = 0$.
Thus we obtain
\begin{equation}
u(x,t + 2K) = z(\alpha - 2L,\beta + 2L) = z(\alpha,\beta) = u(x,t)\>,
\label{u-t-periodicity}
\end{equation}
so that the standing wave solutions are periodic in time, with period
$2K > 2L$.
This implies that the phase (or group) velocity $v = L/K$ is always
smaller than one.
This is consistent with the dispersion relation \eqref{omega-3}
we obtained with the iterative method.

\subsection{Example}
Let us consider an explicit example where we take $b(x) = 0$.
For simplicity, we will choose length units such that $L = \pi$.
Instead of specifying $a(x)$, we take $\pi_t$ of the simple form
\begin{equation}
\pi_t(\lambda) = 1 + A + A\cos \lambda
\end{equation}
for some positive constant $A$.
Using Eq.\ \eqref{pi_t} this implies that
\begin{equation}
a'(\lambda) = \pm 2\sqrt{A}\left|\cos\left(\frac{\lambda}{2}\right)\right|\sqrt{1 + A\cos^2\left(\frac{\lambda}{2}\right)}\>.
\label{example_a'}
\end{equation} 
Now $a'$ satisfies the additional condition
\begin{equation}
\int_0^{2\pi} a'(\lambda) d\lambda = a(2\pi) - a(0) = 0
\end{equation}
because of condition \eqref{bc-periodic}. 
Therefore the sign on the right-hand side of Eq.\ \eqref{example_a'}
has to be taken alternately positive and negative, and we write
\begin{equation}
a'(\lambda) = 2\sqrt{A}\cos\left(\frac{\lambda}{2}\right)
\sqrt{1 + A\cos^2\left(\frac{\lambda}{2}\right)}
\label{example_a'_2}
\end{equation} 
fixing the overall sign to be positive.
Expression \eqref{example_a'_2} can be integrated explicitly, yielding
\begin{equation}
a(\lambda) = 2\sqrt{A\left(1 + \frac{A}{2}\right)}\sin\left(\frac{\lambda}{2}\right)\sqrt{1 + B\cos\lambda}
+ 2(1 + A)\arctan\left(\frac{\sqrt{2B}\sin(\lambda/2)}{\sqrt{1 + B\cos\lambda}}\right)
\label{example_a}
\end{equation}
where we defined $B = A/(2 + A)$.
Note that the integration constant in Eq.\ \eqref{example_a} is fixed
by the antisymmetry condition \eqref{bc-0} on $a$.
For $A \ll 1$, we can approximate
\begin{equation}
a(\lambda) = 4\sqrt{A}\sin\left(\frac{\lambda}{2}\right)
\left[1 + \frac{A}{2}\left(1-\frac{1}{6}\sin^2\left(\frac{\lambda}{2}\right)\right) + \mathcal{O}\bigl(A^2\bigr)\right]\>.
\end{equation}

With these choices for $a(\lambda)$ and $b(\lambda)$
the constant $K$ in Eq.\ \eqref{K} is equal to
$\frac{1}{2}\int_0^{2\pi} \pi_t(\lambda)d\lambda = \pi(1+A)/2$,
so the temporal period of $u(x,t)$ is $2\pi(1 + A)$.
For the coordinate functions $x$ and $t$ we have
\begin{align}
\label{example_x}
x(\alpha,\beta) &= \frac{\alpha + \beta}{2} \\
t(\alpha,\beta) &= (1 + A)\left(\frac{\beta - \alpha}{2}\right) +
A\cos\left(\frac{\alpha + \beta}{2}\right) \sin\left(\frac{\beta - \alpha}{2}\right)
\label{example_t}
\end{align}
We see that in the limit $A \to 0$,
$\alpha = x - t$ and $\beta = x + t$ are lightcone variables,
which shows that, in this limit, $u = z(\alpha, \beta) = a(\alpha) + a(\beta)$
is a superposition of left- and right-moving waves moving at speed 1.
For nonzero $A$, we can write Eq.\ \eqref{example_t} as
\begin{equation}
t = (1 + A)\xi + A\cos x \,\sin \xi
\label{example_t_2}
\end{equation}
where $\xi = (\beta - \alpha)/2$.
It is easy to check that the function $t(\xi)$ defined by Eq.\ \eqref{example_t_2}
is monotonically increasing for any value of $A$ and $x$.
In order to obtain the inverse relation $\xi(t)$ (for any fixed value of $x$)
we define
\begin{equation}
\tau = \frac{t}{1 + A}\>,\qquad\mbox{and}\qquad
\epsilon = \frac{A}{1 + A}\cos x\>.
\label{tau-epsilon}
\end{equation}
It follows that $\xi(t,x) = \lim_{n\to\infty} \xi^{(n)}$,
where $\xi^{(n)}$ is defined recursively by the relation
\begin{equation}
\xi^{(n+1)} = \tau - \epsilon \sin \xi^{(n)}
\end{equation}
with $\xi^{(0)} = \tau$,
so that $\xi$ is given by the infinitely nested expression
\begin{equation}
\xi = \tau - \epsilon \sin(\tau - \epsilon\sin(\tau - \ldots))\>.
\label{example_xi}
\end{equation}
By Taylor-expanding the right-hand side of Eq.\ \eqref{example_xi} around
$\epsilon = 0$ we obtain
\begin{equation}
\xi = \tau - \epsilon\sin \tau + \frac{\epsilon^2}{2}\sin(2\tau)
+ \frac{\epsilon^3}{8}\bigl(\sin \tau - 3\sin(3\tau)\bigr)
+ \frac{\epsilon^4}{6}\bigl(-\sin(2\tau) + 2\sin(4\tau)\bigr)
+ \mathcal{O}(\epsilon^5)\>.
\label{EXAMPLE_XI_2}
\end{equation}
Thus, $\xi$ can be written as a sum of $\tau$ and fluctuations that are
periodic in $\tau$, which can be expressed as a Fourier series,
with coefficients that are expressible as a power series in $\epsilon$.
In Appendix \ref{app:convergence}
it is shown that the series \eqref{EXAMPLE_XI_2} is convergent.

In conclusion, writing $\alpha = x - \xi$, $\beta = x + \xi$,
we can now express 
\begin{equation}
u(x,t) = a(x - \xi) + a(x + \xi)
\end{equation}
with $a$ and $\xi$ given by expressions \eqref{example_a} and
\eqref{example_xi} or \eqref{EXAMPLE_XI_2}, respectively.

\subsection{Standing waves in a constant magnetic field background}

Let us now consider the case of standing waves in a constant magnetic
field background.
To this effect, rather than taking the solution $u(x,t)$ to be periodic,
we search for solutions of the form
\begin{equation}
u(x,t) = B.x + \tilde u(x,t) \>,
\label{u_B}
\end{equation}
with $\tilde u(x,t)$ satisfying the boundary conditions \eqref{boundary-conditions_u}.
The extra term $B.x$ amounts to a constant magnetic field $B_z = B$
(see section \ref{sec:BI-equation}).
For the initial condition we define
\begin{equation}
a(x) = B.x + \tilde a(x)
\label{modified-a}
\end{equation}
with $\tilde{a}(x)$ satisfying the boundary conditions
\eqref{boundary-conditions_a,b}, \eqref{bc-0} and \eqref{bc-periodic}
on $a(x)$.
The conditions on $b(x)$ are unchanged.
Note that $a'(x) = B + \tilde{a}'(x)$ continues to be a
symmetric and periodic function.
It is then easy to check that the properties \eqref{u-antisymmetry},
\eqref{u-x-periodicity} and \eqref{u-t-periodicity} continue to hold,
the temporal periodicity being given by Eq.\ \eqref{K} with the modified
definition \eqref{modified-a} of $a(x)$.
It is worthwhile to point out that in the limit
$|\tilde{a}'(x)|,|b(x)| \ll |B|$, Eq.\ \eqref{K} becomes
\begin{equation}
K \approx L\sqrt{1 + B^2}\>,
\label{K_B-ext-limit}
\end{equation}
so the frequency shift is determined by the background magnetic field.
Thus we find
\begin{equation}
\frac{K}{L} \approx \sqrt{1 + \frac{B^2}{b^2}}\>,
\label{K-over-L}
\end{equation}
where we reinstated the value of the Born-Infeld parameter $b$.
We can compare Eq.\ \eqref{K-over-L} with the dispersion relation
\eqref{dispersion_B-ext-limit} we found using the iterative method,
by identifying $L = 2\pi/k$ and $K = 2\pi/\omega$.
Indeed the expressions match in the limit $B \ll b$. 
However, note that the relation \eqref{K-over-L} is valid even if $B > b$,
as we only need to make sure the hyperbolicity condition
\eqref{hyperbolic_a,b} is satisfied.

Finally, we note that relation \eqref{K-over-L} implies a phase (or group) velocity $v = L/K$ that is once again smaller than one.

\section{Summary and discussion}
\label{sec:discussion}

In this work we considered a class of solution of the Born-Infeld
equation corresponding to standing waves between two parallel 
conducting plates, such that the electric and magnetic field are
parallel to the plates and only depend on time and the coordinate
perpendicular to the plates.
We saw that this amounts to solving the scalar Born-Infeld equation
for the only nonzero component of the vector potential,
a nonlinear partial differential equation in 1+1 dimensions.

We first used the Poincar\'e-Lindstedt iterative method,
which had first been applied by Ferraro \cite{Ferraro} to obtain
an approximate solution to the Born-Infeld equation to order $b^{-2}$,
starting from a seed solution to the linear part of the equation
(which is obtained by taking the $b \to\infty$ limit).
Here we showed how the method can be applied consistently order by order
yielding a solution to the nonlinear differential equation in the form
of a series in terms of inverse powers of the Born-Infeld parameter $b$.

We also applied an alternative ``minimal surface'' method
developed by Barbashov and Chernikov \cite{Barbashov:1966frq}.
It is based on the fact that the 1+1 dimensional scalar Born-Infeld equation
is in fact integrable.
Given initial conditions satisfying hyperbolicity of the partial differential
equation, this procedure yields an exact solution in parametric form.
We showed that, for suitable initial conditions,
standing wave solutions are obtained. 
These solutions satisfy Dirichlet boundary conditions for any time,
and, moreover, they are periodic in time.
The oscillation period was obtained explicitly from the initial conditions.
It is consistent with an effective phase (or group) velocity that is lower
than one.

Using either method, we also studied standing wave solutions in a uniform constant magnetic field background.
Not surprisingly, the nonlinear corrections to the solution,
and also to the dispersion relation will in this case acquire
a dependence on the external magnetic field.
The dispersion relation takes a particularly simple dependence on
the external magnetic field in the limit in which the latter is much larger
than the field amplitude (i.e., the amplitude of the vector potential)
times the wave number
(yielding the magnetic field amplitude of the wave oscillations),
because then the dependence of dispersion relation on the field fluctuations
can be neglected.
In that case there is an interesting comparison we can make with another approach,
that of the ``effective metric'' \cite{Plebanski:1970}
for the propagation of waves in theories of nonlinear electrodynamics
in the presence of background fields.
This formalism has been studied by introducing the so-called Fresnel equation,
which amounts to a dispersion relation for the wave vectors.
It can be derived by studying either the propagation of surfaces of discontinuities
\cite{DeLorenci:2000,Novello:2000,Obukhov:2002,Minz:2014},
or by assuming an approximate a plane-wave ansatz \cite{Schellstede:2015}.
For the case of Born-Infeld electrodynamics,
it follows that the wave vectors $k^\mu$ satisfy the modified dispersion
relation $g^{\mbox{\tiny eff}}_{\mu\nu} k^\mu k^\nu = 0$, where
\begin{equation}
g^{\mbox{\tiny eff}}_{\mu\nu} = \left(1 + \frac{2F}{b^2}\right)\eta_{\mu\nu} + \frac{F_{\mu\lambda}F^\lambda{}_\nu}{b^2}\>.
\label{effective-metric}
\end{equation} 
It is then easy to derive that for the situation studied in this work,
with a background magnetic field $B$ perpendicular to the propagation direction,
the phase velocity of propagation equals
\begin{equation}
v = \frac{k^0}{|\vec{k}|} = \frac{b}{\sqrt{b^2 + B^2}}\>,
\label{v-background}
\end{equation}
in accordance with the (inverse of the) result given by Eq.\ \eqref{K-over-L}.
While this result strictly only applies for the propagation of progressive
waves in a magnetic field background,
we saw in the derivation of the minimal-surface method that we can
consider the standing wave as a superposition of left- and right-moving waves
even in the nonlinear theory.
As the formula \eqref{v-background} applies to either component,
one can expect it to apply as well to the superposition.
In any case, the results in our work validate this interpretation.

Our work also points to interesting experimental possibilities
to test Born-Infeld theory, that is,
to measure or bound the Born-Infeld parameter $b$.
Presumably, the most promising avenue to accomplish this is to consider
a linear resonant cavity and to probe for a frequency dependence,
either on the wave amplitude or on a background magnetic field.
With respect to the latter dependence,
it has been proposed
\cite{Denisov:2004,Zavattini:2008,Grote:2014,Schellstede:2015}
to consider a Michelson interferometer,
and to analyze the effect the application of a background
electromagnetic field in part of one of the legs of the interferometer.
The most promising example would be the interferometers used in the LIGO
or VIRGO experiments to detect gravitational waves,
which work in fact with large resonant cavities inserted in the legs of the
interferometer.
Inside the cavity, the field behaves as a standing wave solution,
rather than a progressive wave.
Our work serves to justify that the $B$ dependence on the frequency is exactly
the same for standing waves as it is for progressive waves.

Our work opens in principle another experimental avenue, 
namely to try and detect the frequency dependence on the field amplitude
itself, given by the relations \eqref{omega-power series} and \eqref{K}.
However, in practice the sensitivity that can be reached this way
with current technology is hampered by the fact that the frequency shifts
that can be obtained this way are much smaller than those with a external background magnetic field, the reason being that even for very strong
laser fields the associated magnetic fields are relatively small.

\acknowledgments

It is our pleasure to thank Claus L\"ammerzahl for suggesting the topic of this work,
and for many illuminating discussions.
Moreover, we are grateful to Robin Tucker for an important comment that triggered us to
write Appendix \ref{app:gauge-choice}.
R.\ P.\ thanks the kind hospitality of the Center for Applied Space Technology
and Microgravity (ZARM) in Bremen, Germany,
and acknowledges financial support by the Funda\c c\~ao para a Ci\^encia e
a Tecnologia of Portugal (FCT) through grant SFRH/BSAB/150324/2019.
Moreover, V.\ P.\ gratefully acknowledges support from  Deutsche
Forschungsgemeinschaft within the Research Training Group  1620 ``Models of Gravity".  

\appendix

\section{Proof that Eq.\ \eqref{gauge-choice} is a good choice of gauge.}
\label{app:gauge-choice}

We consider the subset of field configurations
$\{\vec{E}(\vec{r},t),\vec{B}(\vec{r},t)\}$ satisfying
\begin{align}
E_x(\vec{r},t) &= E_z(\vec{r},t) = 0 \\
E_y(\vec{r},t) &= E_y(x,t) \\
B_x(\vec{r},t) &= B_y(\vec{r},t) = 0 \\
B_z(\vec{r},t) &= B_z(x,t)\>.
\end{align}
We will show that for such configurations, Eq.\ (8) is a good choice of gauge.
To see this, we first apply a gauge transformation such that we obtain the
temporal gauge:
\begin{equation}
\Phi(\vec{r},t) = A_0(\vec{r},t) = 0\>.
\label{temporal-gauge}
\end{equation}
We then have
\begin{align}
\label{dt-Ax}
E_x(\vec{r},t) &= -\partial_t A_x = 0 \\
E_y(\vec{r},t) &= -\partial_t A_y =  E_y(x,t) \\
\label{dt-Az}
E_z(\vec{r},t) &= -\partial_t A_z = 0 \\
B_x(\vec{r},t) &= \partial_y A_z - \partial_z A_y = 0 \\
\label{By}
B_y(\vec{r},t) &= \partial_z A_x - \partial_x A_z = 0 \\
\label{Bz}
B_z(\vec{r},t) &= \partial_x A_y - \partial_y A_x = B_z(x,t)\>.
\end{align}
We see from Eqs.\ \eqref{dt-Ax} and \eqref{dt-Az} that $A_x$ and $A_z$ are
time-independent.
Now note that the gauge condition \eqref{temporal-gauge} still allows a
residual gauge transformation
\begin{equation}
\vec{A} \to \vec{A} + \vec{\nabla}\epsilon_1(\vec{r})
\label{residual-gauge-transformation1}
\end{equation}
for arbitrary time-independent gauge parameter $\epsilon_1(\vec{r})$.
We can use this freedom to fix the additional gauge condition
\begin{equation}
A_x(\vec{r}) = 0\>.
\label{additional-gauge-condition1}
\end{equation}
Note that conditions \eqref{temporal-gauge} and \eqref{additional-gauge-condition1}
still leave another residual gauge invariance
\begin{equation}
\vec{A} \to \vec{A} + \vec{\nabla}\epsilon_2(y,z)
\label{residual-gauge-transformation2}
\end{equation}
for some arbitrary gauge parameter $\epsilon_2(y,z)$.
Now note that it follows from Eqs.\ \eqref{dt-Az}, \eqref{By} and \eqref{additional-gauge-condition1}
that $A_z$ only depends on the coordinates $y$ and $z$.
We can use the gauge freedom \eqref{residual-gauge-transformation2} to fix the
additional gauge condition to eliminate $A_z$:
\begin{equation}
A_z(y,z) = 0\>.
\label{additional-gauge-condition2}
\end{equation}
Eqs.\  \eqref{dt-Ax}--\eqref{Bz} then reduce to
\begin{align}
\label{dt-Ay}
-\partial_t A_y(\vec{r},t) &= E_y(x,t)\\
\partial_x A_y(\vec{r},t) &= B_z(x,t)\\
\partial_z A_y(\vec{r},t) &= 0\>.
\label{dz-Ay}
\end{align}
Conditions \eqref{dt-Ay}--\eqref{dz-Ay} fix the gauge component $A_y$ up to an
arbitrary function $f(y)$ of the coordinate $y$ only.
Note, however, that gauge conditions \eqref{temporal-gauge},
\eqref{additional-gauge-condition1} and \eqref{additional-gauge-condition2}
still allow for a (third) residual gauge invariance 
\begin{equation}
\vec{A} \to \vec{A} + \vec{\nabla}\epsilon_3(y)\>.
\label{residual-gauge-transformation3}
\end{equation}
We can use this freedom to set $f(y) = 0$,
so that $A_y$ only depends on $t$ and $x$.
We are left with the final relations
\begin{align}
-\partial_t A_y(x,t) &= E_y(x,t)\\
\partial_x A_y(x,t) &= B_z(x,t)\>.
\end{align}

\section{Properties of the asymptotic series solution of Section \ref{sec:iterative}}
\label{app:asymptotic}

In this appendix we will explicitly work out the recursion formulas for 
constructing the asymptotic series solution of Section \ref{sec:iterative} 
and we will numerically estimate how accurately it satisfies 
the Born-Infeld  equation when cut off after a few terms. 

\begin{figure}[h]
\begin{center} 
\includegraphics[width=14cm]{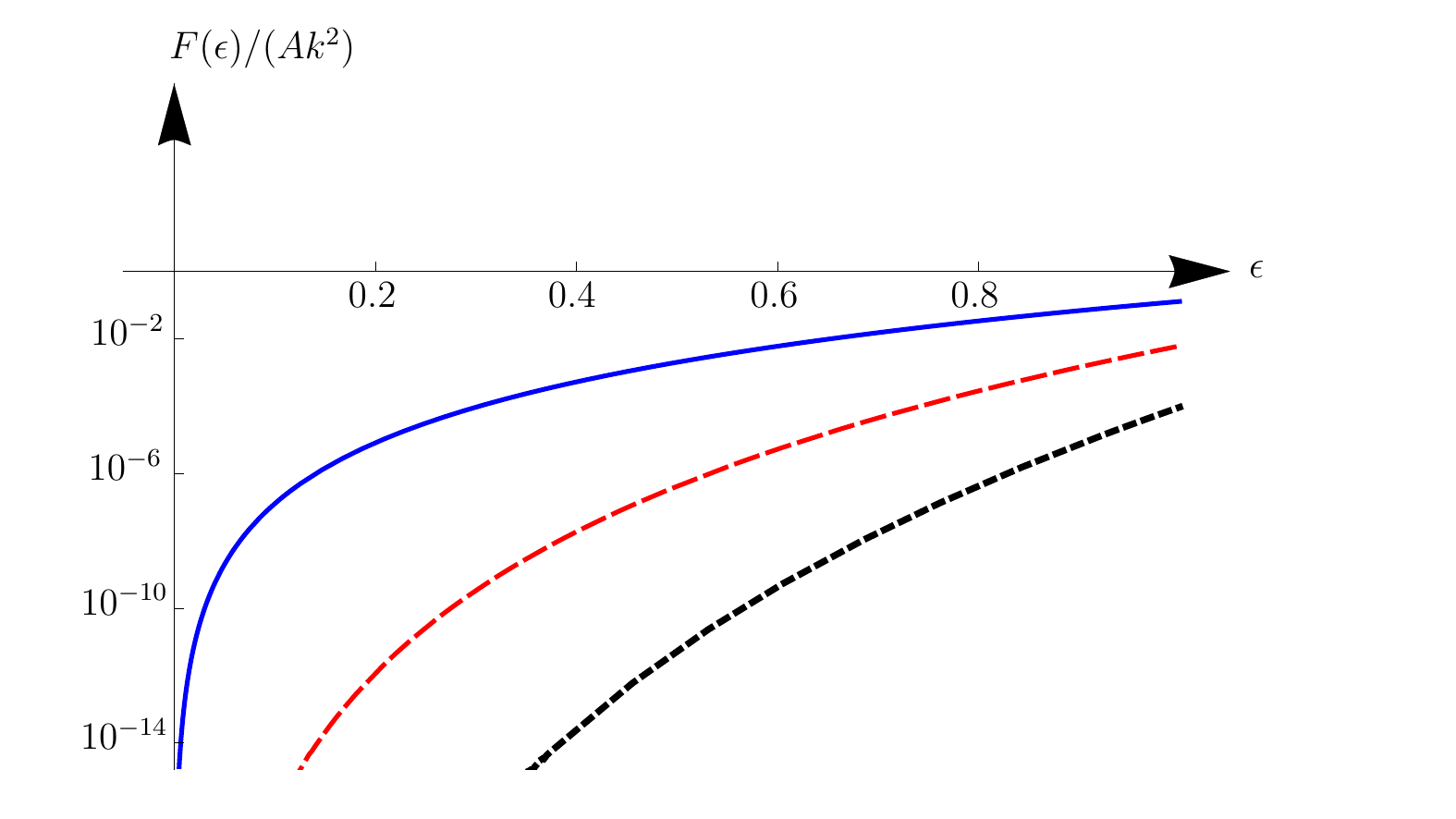}    
\end{center} 
\caption{\small
Error by which the $N$th order asymptotic series fails to
satisfy the Born-Infeld equation, for $N=3$ (solid), $N=6$
(dashed) and $N=11$ (dotted). $F( \epsilon )$ denotes 
the maximum over $x$ and $t$ of the left-hand side of 
the Born-Infeld equation (\protect\ref{BI-equation-b}). 
Note that $F ( \epsilon ) / (Ak^2)$
is dimensionless.}  \label{fig:error}
\end{figure}

To derive the recursion formulas, we start out from the seed 
solution (\ref{eq:seed}), thereby fixing
an amplitude $A$ and a wave number $k$. During the entire
procedure $A$ and $k$ will be kept fixed. 
The ansatz for the $N$th order solution is 
\begin{equation}
u  (x,t) =
A  \sum _{M=0} ^N 
\sum _{\nu =0} ^M \sum _{\mu =0} ^M 
\alpha _{M \nu \mu} \,
\mathrm{sin} \Big( (2 \nu + 1 ) kx \Big) \, \mathrm{cos} \Big( ( 2 \mu + 1 )  \omega _N t \Big) \, 
\epsilon ^{2M}
\label{eq:useries}
\end{equation}
where the frequency $\omega _N$ depends on $\epsilon = Ak/b$ according to 
\begin{equation}
\omega _N  ^2 = k^2 \sum _{M=0} ^N \xi_M \epsilon ^{2M} \, .
\end{equation}
It is our goal to derive recursion formulas for the coefficients 
$\alpha _{M \nu \mu}$ and $\xi _M$ which have to satisfy   
\begin{equation}
\alpha _{M \nu \mu} = 0 \quad \text{if} \: \, M \ge 1 \: \text{and} \; \mu = \nu \, , \quad
\alpha _{000}= 1 \, , \quad \xi _0 = 1 \, .
\end{equation}
Inserting ansatz (\ref{eq:useries}) into the Born-Infeld equation (\ref{BI-equation-b})
and dividing by $A k^2$ results in
\[
0 
= 
\sum _{M=0} ^N \sum _{\nu=0} ^M \sum _{\mu=0} ^M 
\alpha _{M \nu \mu} \,
\Bigg( (2 \mu +1)^2  \sum _{\hat{M}=0} ^N \xi _{\hat{M}}  \epsilon ^{2 \hat{M}} -(2 \nu + 1 )^2 \Bigg)
\mathrm{sin} \Big( (2 \nu + 1 ) kx \Big) \, \mathrm{cos} \Big( (2 \mu +1)  \omega _N t \Big) \, 
\epsilon ^{2M}
\]
\[
+ \, \epsilon ^2 
\sum _{\hat{M}=0} ^N  
\sum _{\tilde{M}=0} ^N \sum _{\tilde{\nu}=0} ^{\tilde{M}} \sum _{\tilde{\mu}=0} ^{\tilde{M}}
\sum _{M'=0} ^N \sum _{\nu '=0} ^{M'} \sum _{\mu '=1} ^{M'}
\sum _{M''=0} ^N \sum _{\nu ''=0} ^{M''} \sum _{\mu ''=0} ^{M''}
\xi _{\hat{M}} \, \alpha _{\tilde{M} \tilde{\nu} \tilde{\mu}} \, \alpha _{M' \nu ' \mu '} \, \alpha _{M'' \nu '' \mu ''}
\, \epsilon ^{2(\tilde{M}+M'+M'' +\hat{M})}
\]
\[
\times \Bigg(
(1 + 2 \mu ') (1 + 2 \mu '')  (1 + 2 \tilde{\nu} )^2   
\mathrm{sin} \Big( (1 + 2 \nu ') kx \Big) 
\mathrm{sin} \Big( (1 + 2 \nu '') kx \Big) 
\mathrm{sin} \Big( (1 + 2 \tilde{\nu} ) kx \Big) 
\]
\[
\times
\mathrm{sin} \Big( (1 + 2 \mu ') \omega _N t  \Big) 
\mathrm{sin} \Big( (1 + 2 \mu '')  \omega _N t  \Big) 
\mathrm{cos} \Big( (1 + 2 \tilde{\mu})  \omega _N t  \Big) 
\Bigg.
\]
\[
+
(1 + 2 \nu ') (1 + 2 \nu '')  (1 + 2 \tilde{\mu} )^2   
\mathrm{cos} \Big( (1 + 2 \nu ') kx \Big) 
\mathrm{cos} \Big( (1 + 2 \nu '') kx \Big) 
\mathrm{sin} \Big( (1 + 2 \tilde{\nu} ) kx \Big) 
\]
\[
\times
\mathrm{cos} \Big( (1 + 2 \mu ') \omega _N t  \Big) 
\mathrm{cos} \Big( (1 + 2 \mu '')  \omega _N t  \Big) 
\mathrm{cos} \Big( (1 + 2 \tilde{\mu})  \omega _N t  \Big) 
\]
\[
\Bigg.
+ 2
(1 + 2 \mu ') (1 + 2 \nu '')  (1 + 2 \tilde{\nu} ) (1 + 2 \tilde{\mu} )   
\mathrm{sin} \Big( (1 + 2 \nu ') kx \Big) 
\mathrm{cos} \Big( (1 + 2 \nu '') kx \Big) 
\mathrm{cos} \Big( (1 + 2 \tilde{\nu} ) kx \Big) 
\]
\begin{equation}
\times
\mathrm{sin} \Big( (1 + 2 \mu ') \omega _N t  \Big) 
\mathrm{cos} \Big( (1 + 2 \mu '')  \omega _N t  \Big) 
\mathrm{sin} \Big( (1 + 2 \tilde{\mu})  \omega _N t  \Big) 
\Bigg)
\label{eq:BIseries}
\end{equation}
For the triple products of sine and cosine functions we use trigonometric 
identities, e.g.  
\begin{equation}
4 \, \mathrm{sin} \, \alpha \, \mathrm{sin} \, \beta \, \mathrm{sin} \gamma =
- \mathrm{sin} \big( \alpha - \beta - \gamma \big) 
+ \mathrm{sin} \big( \alpha + \beta - \gamma \big) 
+ \mathrm{sin} \big( \alpha - \beta + \gamma \big) 
- \mathrm{sin} \big( \alpha + \beta + \gamma \big) 
\, .
\end{equation}
In this way we rewrite the right-hand side of Eq.\ (\ref{eq:BIseries}) as a sum over terms
proportional to $\mathrm{sin} \big( (2 \nu + 1 ) kx \big) \, 
\mathrm{cos} \big( (2 \mu + 1 ) \omega _N t \big) \epsilon ^{2M}$
which are linearly independent for different values of $(M, \nu , \mu)$.
Equating to zero the coefficients of 
$\mathrm{sin} \big((2 \nu + 1 ) kx \big) \, 
\mathrm{cos} \big( (2 \mu + 1 ) \omega _N t \big) \epsilon ^{2N}$ 
for $N \ge 1$ and $(\nu, \mu ) \neq (0,0)$ results in
\[
0 
= \Big( (2 \mu + 1 ) ^2 - (2 \nu + 1 ) ^2 \Big) \alpha _{N \nu \mu}
+ 
\sum _{M=1} ^{N-1}
\alpha _{M \nu \mu} \,
 (2 \mu +1)^2  \xi _{N-M} 
\]
\[
+ \, \dfrac{1}{16} 
\sum _{\tilde{M}=0} ^{N-1} \sum _{\tilde{\nu}=0} ^{\tilde{M}} \sum _{\tilde{\mu}=0} ^{\tilde{M}}
\sum _{M'=0} ^{N-1} \sum _{\nu '=0} ^{M'} \sum _{\mu '=1} ^{M'}
\sum _{M''=0} ^{N-1} \sum _{\nu ''=0} ^{M''} \sum _{\mu ''=0} ^{M''}
\xi _{N- M' - M'' - \tilde{M}-1} \, \, \alpha _{M' \nu ' \mu '} \, \alpha _{M'' \nu '' \mu ''}
\alpha _{\tilde{M} \tilde{\nu} \tilde{\mu}} 
\]
\[
\times 
\Big(
(2 \mu ' +1) ( 2 \mu '' + 1 ) (2 \tilde{\nu} + 1 )^2 
Q_1 \big( \nu , \nu ' , \nu '' , \tilde{\nu} \big)
P_1 \big( \mu , \mu ' , \mu '' , \tilde{\mu} \big)
\Big.
\]
\[
+
(2 \nu ' + 1) (2 \nu '' + 1 ) (2 \tilde{\mu} + 1 )^2 
Q_2 \big( \nu , \nu ' , \nu '' , \tilde{\nu} \big)
P_2 \big( \mu , \mu ' , \mu '' , \tilde{\mu} \big)
\]
\begin{equation}\label{eq:alpha}
\Big.
+
(2 (2 \mu ' +1 ) (2 \nu '' + 1 ) ( 2 \tilde{\nu} + 1 )  ( 2 \tilde{\mu} + 1 )  
Q_3 \big( \nu , \nu ' , \nu '' , \tilde{\nu} \big)
P_3 \big( \mu , \mu ' , \mu '' , \tilde{\mu} \big)
\Big)
\end{equation}
where
\[
Q_1 \big( \nu , \nu ' , \nu '' , \tilde{\nu} \big)
=
\delta ^{\tilde{\nu}} _{ -2 - \nu - \nu ' - \nu ''}
- 
\delta ^{\tilde{\nu}} _{  - 1 + \nu - \nu ' - \nu ''}
-
\delta ^{\tilde{\nu}} _{ -1 -  \nu + \nu ' - \nu ''}
+
\delta ^{\tilde{\nu}} _{ \nu +\nu ' - \nu ''}
\]
\begin{equation}
-
\delta ^{\tilde{\nu}} _{-1 -   \nu - \nu ' + \nu ''}
+
\delta ^{\tilde{\nu}} _{ \nu -\nu ' + \nu ''}
+
\delta ^{\tilde{\nu}} _{ - \nu + \nu ' + \nu ''}
-
\delta ^{\tilde{\nu}} _{1 + \nu + \nu ' + \nu ''}
\, ,
\end{equation}
\[
Q_2 \big( \nu , \nu ' , \nu '' , \tilde{\nu} \big)
=
-\delta ^{\tilde{\nu}} _{ -2 - \nu - \nu ' - \nu ''}
+
\delta ^{\tilde{\nu}} _{  - 1 + \nu - \nu ' - \nu ''}
-
\delta ^{\tilde{\nu}} _{ -1 -  \nu + \nu ' - \nu ''}
+
\delta ^{\tilde{\nu}} _{ \nu +\nu ' - \nu ''}
\]
\begin{equation}
-
\delta ^{\tilde{\nu}} _{-1 -   \nu - \nu ' + \nu ''}
+
\delta ^{\tilde{\nu}} _{ \nu -\nu ' + \nu ''}
-
\delta ^{\tilde{\nu}} _{ - \nu + \nu ' + \nu ''}
+
\delta ^{\tilde{\nu}} _{1 + \nu + \nu ' + \nu ''}
\, ,
\end{equation}
\[
Q_3 \big( \nu , \nu ' , \nu '' , \tilde{\nu} \big)
=
\delta ^{\tilde{\nu}} _{ -1 + \nu - \nu ' - \nu ''}
- 
\delta ^{\tilde{\nu}} _{  \nu + \nu ' - \nu ''}
+
\delta ^{\tilde{\nu}} _{  \nu - \nu ' + \nu ''}
-
\delta ^{\tilde{\nu}} _{1+  \nu +\nu ' + \nu ''}
\]
\begin{equation}
+
\delta ^{\tilde{\nu}} _{-  \nu +\nu ' + \nu ''}
-
\delta ^{\tilde{\nu}} _{- 1 - \nu -\nu ' + \nu ''}
+
\delta ^{\tilde{\nu}} _{- 1 - \nu + \nu ' - \nu ''}
-
\delta ^{\tilde{\nu}} _{- 2 - \nu - \nu ' - \nu ''}
\, ,
\end{equation}
\[
P_1 \big( \mu , \mu ' , \mu '' , \tilde{\mu} \big)
=
-\delta ^{\tilde{\mu}} _{ -2 - \mu - \mu ' - \mu ''}
- 
\delta ^{\tilde{\mu}} _{ -1+ \mu - \mu ' - \mu ''}
+
\delta ^{\tilde{\mu}} _{ -1 - \mu + \mu ' - \mu ''}
+
\delta ^{\tilde{\mu}} _{\mu +\mu ' - \mu ''}
\]
\begin{equation}
+
\delta ^{\tilde{\mu}} _{-1-  \mu - \mu ' + \mu ''}
+
\delta ^{\tilde{\mu}} _{ \mu -\mu ' + \mu ''}
-
\delta ^{\tilde{\mu}} _{ - \mu + \mu ' + \mu ''}
-
\delta ^{\tilde{\mu}} _{1 + \mu + \mu ' + \mu ''}
\, ,
\end{equation}
\[
P_2 \big( \mu , \mu ' , \mu '' , \tilde{\mu} \big)
=
\delta ^{\tilde{\mu}} _{ -2 + \mu - \mu ' - \mu ''}
+ 
\delta ^{\tilde{\mu}} _{ \mu + \mu ' - \mu ''}
+
\delta ^{\tilde{\mu}} _{  \mu - \mu ' + \mu ''}
+
\delta ^{\tilde{\mu}} _{1+ \mu +\mu ' + \mu ''}
\]
\begin{equation}
+
\delta ^{\tilde{\mu}} _{-  \mu + \mu ' + \mu ''}
+
\delta ^{\tilde{\mu}} _{-1 -  \mu -\mu ' + \mu ''}
+
\delta ^{\tilde{\mu}} _{ -1 - \mu + \mu ' - \mu ''}
+
\delta ^{\tilde{\mu}} _{-2 - \mu - \mu ' - \mu ''}
\, ,
\end{equation}
\[
P_3 \big( \mu , \mu ' , \mu '' , \tilde{\mu} \big)
=
-\delta ^{\tilde{\mu}} _{ -2 - \mu - \mu ' - \mu ''}
- 
\delta ^{\tilde{\mu}} _{ -1+ \mu - \mu ' - \mu ''}
+
\delta ^{\tilde{\mu}} _{ -1 - \mu + \mu ' - \mu ''}
+
\delta ^{\tilde{\mu}} _{\mu +\mu ' - \mu ''}
\]
\begin{equation}
-
\delta ^{\tilde{\mu}} _{-1-  \mu - \mu ' + \mu ''}
-
\delta ^{\tilde{\mu}} _{ \mu -\mu ' + \mu ''}
+
\delta ^{\tilde{\mu}} _{ - \mu + \mu ' + \mu ''}
+
\delta ^{\tilde{\mu}} _{1 + \mu + \mu ' + \mu ''}
\, ,
\end{equation}
with $\delta ^{\sigma} _{\tau}$ denoting the Kronecker delta.
Eq.\ (\ref{eq:alpha}) determines $\alpha _{N \nu \mu}$ in terms of the 
lower-order coeffcients, $\alpha _{M \nu \mu}$ and $\xi _{M}$
for $M \le N-1$. Similarly, equating to zero the coefficients of 
$\mathrm{sin} \big(kx \big) \, 
\mathrm{cos} \big( \omega _N t \big) \epsilon ^{2N}$ results in
\[
0 
= \xi _N 
+ \dfrac{1}{16} 
\sum _{\tilde{M}=0} ^{N-1} \sum _{\tilde{\nu}=0} ^{\tilde{M}} \sum _{\tilde{\mu}=0} ^{\tilde{M}}
\sum _{M'=0} ^{N-1} \sum _{\nu '=0} ^{M'} \sum _{\mu '=1} ^{M'}
\sum _{M''=0} ^{N-1} \sum _{\nu ''=0} ^{M''} \sum _{\mu ''=0} ^{M''}
\xi _{N- M' - M'' - \tilde{M}-1} \, \, \alpha _{M' \nu ' \mu '} \, \alpha _{M'' \nu '' \mu ''}
\alpha _{\tilde{M} \tilde{\nu} \tilde{\mu}} 
\]
\[
\times
\Big(
(2 \mu ' +1) ( 2 \mu '' + 1 ) (2 \tilde{\nu} + 1 )^2 
Q_1 \big( 0 , \nu ' , \nu '' , \tilde{\nu} \big)
P_1 \big( 0 , \mu ' , \mu '' , \tilde{\mu} \big)
\Big.
\]
\[
+
(2 \nu ' + 1) (2 \nu '' + 1 ) (2 \tilde{\mu} + 1 )^2 
Q_2 \big( 0 , \nu ' , \nu '' , \tilde{\nu} \big)
P_2 \big( 0 , \mu ' , \mu '' , \tilde{\mu} \big)
\]
\begin{equation}\label{eq:xi}
\Big.
+
(2 (2 \mu ' +1 ) (2 \nu '' + 1 ) ( 2 \tilde{\nu} + 1 )  ( 2 \tilde{\mu} + 1 )  
Q_3 \big( 0 , \nu ' , \nu '' , \tilde{\nu} \big)
P_3 \big( 0 , \mu ' , \mu '' , \tilde{\mu} \big)
\Big)
\end{equation}
Eq.\ (\ref{eq:xi})  determines $\xi _N$ in terms of the lower-order
coefficients, $\xi _M$ and $\alpha _{M \nu \mu}$ for $M\le N-1$.
The recursive formulas (\ref{eq:alpha}) and (\ref{eq:xi}) demonstrate
that the coefficients of our series are indeed well-defined for every $N$. 
Moreover, they provide bounds on the coefficients of order $N$ in terms 
of the coefficients of order $M \le N-1$. However, we have 
not been able to prove, with the help of such bounds, that the series 
converges (pointwise or in any other sense) for $N \to \infty$. In any 
case, it is an asymptotic series for 
$\epsilon \to 0$. To estimate how accurately the $N$th order
solution satisfies the Born-Infeld equation, for low $N$, we have 
numerically calculated the coefficients $\alpha _{M \nu \mu}$
and $\xi _M$ up to $M=11$ and inserted the resulting
$N$th order series solution into the left hand-side of the
Born-Infeld equation (\ref{BI-equation-b}). We have then plotted
the maximum over $x$ and $t$ of this function against $\epsilon$,
for $N=3$, $N=6$ and $N=11$, see Fig.~\ref{fig:error}. 
Note the logarithmic scale on the vertical axis. 
Recall that $\epsilon = Ak/b$ is certainly very small
for all electromagnetic fields that have been produced in 
the laboratory so far, because otherwise deviations from 
the standard Maxwell theory would have been observed 
already. Therefore, we may safely assume that in the 
foreseeable future all experimental tests of the Born-Infeld 
theory will be performed with fields for which $\epsilon < 0.1$, 
say. We read from Fig.~\ref{fig:error} that in this regime 
our asymptotic series solutions satisfy the Born-Infeld 
equation with a high accuracy, even if cut off after a few 
terms.     

We have mentioned already that we conjecture that the
series actually converges for $N \to \infty$ towards an exact 
solution. Our numerical results certainly give some hope that 
this conjecture is true, but we do not have a proof. There
are examples of asymptotic series where the summands
go down very rapidly up to a certain order and then
begin to increase again. We cannot rule out that this is
also the case for our series.

\section{Convergence of the Fourier series \eqref{EXAMPLE_XI_2}  }
\label{app:convergence}

It can be verified by inspection
that the periodic functions in $\tau$ 
on the right-hand side of Eq.\ \eqref{EXAMPLE_XI_2} multiplying
the succesive powers of $\epsilon$ are bounded by 1 in absolute value
for any value of the argument $\tau$.
Therefore, the series should be absolutely convergent for $|\epsilon| < 1$.
With the identification \eqref{tau-epsilon},
this means we expect the series \eqref{EXAMPLE_XI_2} to be convergent
for any value of $A$ and $x$.

There is another,
rigorous argument that shows convergence of the Fourier series
\eqref{EXAMPLE_XI_2}.
Relation \eqref{example_t_2}
\begin{equation}
\tau(\xi,\epsilon) = \xi + \epsilon \sin\xi\>,
\label{tau-xi-epsilon}
\end{equation}
(with $\tau$ and $\epsilon$ defined by Eqs.\ \eqref{tau-epsilon})
can be considered as an analytic function $\tau$ of complex variables $\xi$
and $\epsilon$.
We already argued relation \eqref{tau-xi-epsilon} can be inverted to yield
$\xi$ as a function of $\tau$ for any real values of $\tau$,
as long as $|\epsilon| < 1$.
Noting that $\tau$ is an analytic function of $\xi$ and $\epsilon$,
the same should be true for the inverse function $\xi(\tau,\epsilon)$,
at least in a subdomain of $\mathbb{C}\otimes\mathbb{C}$ that includes
the subset $\mathbb{R}\otimes\, ]{-1},1[$.
Writing $\xi = \xi_1 + i\xi_2$, $\tau = \tau_1 + i\tau_2$,
$\epsilon = \epsilon_1 + i\epsilon_2$ as the sum of real and imaginary
parts, Eq.\ \eqref{tau-xi-epsilon} yields the equtions
\begin{align}
\label{tau-xi-epsilon-Re}
\tau_1 &= \xi_1 + \epsilon_1\sin\xi_1 \cosh\xi_2 - \epsilon_2 \cos\xi_1 \sinh\xi_2 \\
\tau_2 &= \xi_2 + \epsilon_2 \sin\xi_1 \cosh\xi_2 + \epsilon_1 \cos\xi_1 \sinh\xi_2
\label{tau-xi-epsilon-Im}
\end{align}
Taking $\tau$ real ($\tau_2 = 0$), and representing
$\epsilon_1 = |\epsilon|\cos\phi$,  $\epsilon_2 = |\epsilon|\sin\phi$,
Eq.\ \eqref{tau-epsilon} becomes
\begin{equation}
-\xi_2 = |\epsilon|(\cos\phi \cos\xi_1 \sinh\xi_2 + \sin\phi \sin\xi_1 \cosh\xi_2) \>.
\label{tau-xi-epsilon-Im2}
\end{equation}
Calling the right-hand side $f(\xi_2)$, it is easy to see that
$f(\xi_2) \le |\epsilon| \cosh\xi_2$ for any $\xi_1$ and $\phi$.
Thus, the graph of the function $f(\xi_2)$ lies below that of
$\epsilon\cosh\xi_2$.
Considering now the graphs of the functions $-\xi_2$ and $\epsilon\cosh\xi_2$, Eq.\ \eqref{tau-xi-epsilon-Im2} is guaranteed to have a solution
for $\xi_2$ for arbitrary given values of $\phi$ and $\xi_1$,
as long as the graphs of the functions $-\xi_2$ and $\epsilon\cosh\xi_2$
intersect.
It is easy to see that this is the case for $|\epsilon| \le \epsilon_c$,
where the critical value $\epsilon_c$ is the maximum value of $\epsilon$
such that the equation $x = \epsilon \cosh x$ has at least one (real) root.
One finds $\epsilon_c \approx 0.663$.
It also follows that the absolute value of at least one of the solutions
of Eq.\ \eqref{tau-xi-epsilon-Im2} for $\xi_2$ is smaller than
$(\arcsinh \epsilon_c)^{-1} \approx 1.19962$.

Once a solution is found for $\xi_2$ of Eq.\ \eqref{tau-xi-epsilon-Im2},
it can be seen that there exists at least one solution of
Eq.\ \eqref{tau-xi-epsilon-Re} for $\xi_1$, if we take the above-mentioned
minimal bounded solution for $\xi_2$. 
Namely, if $\xi_2$ is bounded, the same will be true for the last two terms
on the right-hand side of Eq.\ \eqref{tau-xi-epsilon-Re}.
Therefore, it is always possible to find a value of $\xi_1$ such that
the sum of the terms on the right-hand side of Eq.\ \eqref{tau-xi-epsilon-Re}
equals any given value of $\tau_1$. 
 
From the above we can conclude that Eq.\ \eqref{tau-xi-epsilon} can be
inverted to yield $\xi(\tau,\epsilon)$, for any real value of $\tau$
and any complex $\epsilon$ such that $|\epsilon| \le \epsilon_c$.
This means in turn that, for any fixed value of $\tau$,
the analytic function $\xi(\tau,\epsilon)$ can be expanded in a
power series in the complex variable $\epsilon$, which converges to 
$\xi(\tau,\epsilon)$ at least for $|\epsilon| \le \epsilon_c$.

\end{document}